\documentclass[aps,prb,reprint,letterpaper]{revtex4-2}
\pdfoutput=1
\include{math}
\usepackage{bm}
\usepackage{graphicx}
\usepackage{amsmath}
\usepackage{booktabs}
\usepackage{natbib}
\usepackage{url}
\usepackage{textcase}

\begin{document}

\title{Chiral properties of the zero-field spiral state and field-induced magnetic phases of the itinerant kagome metal YMn$_6$Sn$_6$}

\author{Rebecca L. Dally}
\email[corresponding author: ]{rebecca.dally@nist.gov}
\affiliation{NIST Center for Neutron Research, National Institute of Standards and Technology, Gaithersburg, MD 20899-6102}
\author{Nirmal J. Ghimire}
\affiliation{Department of Physics and Astronomy, George Mason University, Fairfax, VA 22030}
\affiliation{Quantum Science and Engineering Center, George Mason University, Fairfax, VA 22030}
\author{Dina Michel}
\affiliation{Department of Physics and Astronomy, George Mason University, Fairfax, VA 22030}
\affiliation{Quantum Science and Engineering Center, George Mason University, Fairfax, VA 22030}
\author{Peter Siegfried}
\affiliation{Department of Physics and Astronomy, George Mason University, Fairfax, VA 22030}
\affiliation{Quantum Science and Engineering Center, George Mason University, Fairfax, VA 22030}
\author{Igor I. Mazin}
\affiliation{Department of Physics and Astronomy, George Mason University, Fairfax, VA 22030}
\affiliation{Quantum Science and Engineering Center, George Mason University, Fairfax, VA 22030}
\author{Jeffrey W. Lynn}
\affiliation{NIST Center for Neutron Research, National Institute of Standards and Technology, Gaithersburg, MD 20899-6102}

\date{\today}
\begin{abstract}

Applying a magnetic field in the hexagonal plane of YMn$_6$Sn$_6$ leads to a complex magnetic phase diagram of commensurate and incommensurate phases, one of which coexists with the topological Hall effect (THE) generated by a unique fluctuation-driven mechanism. Using unpolarized neutron diffraction, we report on the solved magnetic structure for two previously identified, but unknown, commensurate phases. These include a low-temperature, high-field fan-like phase and a room-temperature, low-field canted antiferromagnetic phase. An intermediate incommensurate phase between the fan-like and forced ferromagnetic phases is also identified as the last known phase of the in-plane field-temperature diagram. Additional characterization using synchrotron powder diffraction reveals extremely high-quality, single-phase crystals, which suggests that the presence of two incommensurate magnetic structures throughout much of the phase diagram is an intrinsic property of the system. Interestingly, polarized neutron diffraction shows that the centrosymmetric system hosts preferential chirality in the zero-field double-flat-spiral phase, which, along with the THE, is a topologically non-trivial characteristic.    

\end{abstract}
\maketitle

\section{Introduction}
Verifying the correct ground state for magnetic systems with competing interactions has been a fundamental problem since the triangular Ising antiferromagnet was first studied 70 years ago. This is an example of antiferromagnetic (AF) interactions on a particular lattice geometry that leads to magnetic frustration, \textit{e.g.} geometrical frustration. Moving beyond frustration due solely to geometrical restrictions combined with AF interactions, one can look to competing nearest neighbor and next-nearest neighbor -- and farther -- interactions, which can lead to either no order, as is the case in spin liquids, short-range order, or even a multi-phase space, where either side of a phase boundary line represents two different orderings with subtle energetic differences. Often, the structure which emerges from the frustration is a long-wavelength incommensurate spin texture, where the details of the underlying crystal lattice symmetry determine additional expressed features, \cite{tokura2018nonreciprocal} such as chiral handedness, \cite{grohol2005, Zorko_PRL_2011} the magnetoelectric effect, \cite{Agyei_1990, Edelstein_PRL_1995}, toroidal order, \cite{Lei2020} and non-reciprocal magnons.~\cite{Gitgeatpong_PRL_2017, Weber_AIP_2018} These are examples of phenomena which occur when magnetism is in the presence of broken spatial inversion symmetry (\textit{i.e.} non-centrosymmetric lattices). 

More recently, magnetic frustration in centrosymmetric systems has been theorized, and experimentally verified, as a route to stabilize topologically protected skyrmion lattices, \cite{Kurumaji914, OkuboPRL2012, leonov2015NC} a phase that traditionally materialized from chiral crystal structures. Similarly, topologically non-trivial multi-$\mathbf{q}$ structures, other than the canonical triple-$\mathbf{q}$ of the skyrmion lattice, and in the absence of the Dzyaloshinskii-Moriya antisymmetric exchange interaction, have also been reported.~\cite{Ishiwata_PRB_2020} It is then natural to ask whether other topological properties that require broken inversion symmetry can be found in centrosymmetric, frustrated magnet systems. For example, the topological Hall effect (THE) was recently observed in YMn$_6$Sn$_6$, \cite{wang2019_the, ghimire2020} a centrosymmetric (space group $P6/mmm$) itinerant helimagnet, with the maximum effect occurring around 245 K and an applied field of about 4 T in the $ab$-plane. Although no skyrmion lattice was found in this region of phase space, a non-coplanar spin texture was: a transverse conical spiral.~\cite{ghimire2020, neubauer2020plane} This spin texture would not on its own lead to the THE, but it was argued that dynamic chiral fluctuations are responsible, thus making YMn$_6$Sn$_6$ a prototype material for a fluctuation based THE mechanism. Thermal fluctuations, coupled with the strongly two-dimensional nature of the magnetic exchange, are thought to be key ingredients for realizing the THE despite the null scalar spin chirality in the absence of an external field. It is then the addition of unbalanced magnon fluctuations in the transverse conical phase which creates a nonzero chiral susceptibility.  

\begin{figure}[t]
\includegraphics[scale=0.155]{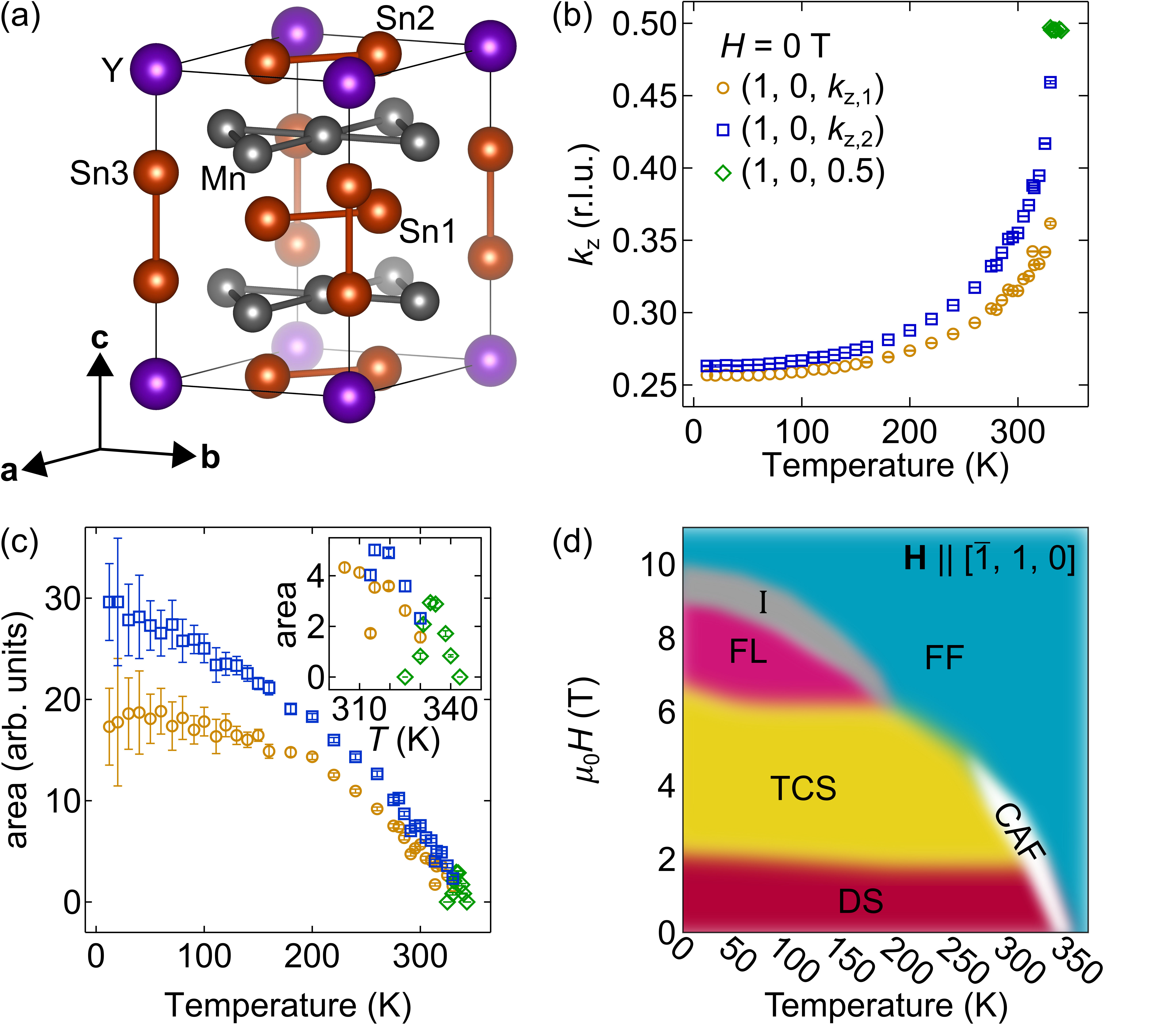}
\caption{(a) Crystal structure of YMn$_6$Sn$_6$, belonging to space group $P6/mmm$ (191). Single-crystal neutron diffraction data showing the temperature dependence of the magnetic (b) wavevectors and (c) Bragg peak intensities in zero-field conditions. The inset of (c) shows the progression of the magnetic structures just below $T_N \approx 340$ K. One commensurate structure emerges at $T_N$, but it is short lived and quickly gives way with decreasing temperature to two incommensurate structures. The legend for (c) is the same as in (b). (d) Magnetic structure phase diagram for the applied field, $\mathbf{H}$, in-plane. The neutron diffraction experiments in this report were carried out with $\mathbf{H} \parallel [\bar{1}, 1, 0]$. The phases DS, TCS, FL, CAF, I, and FF correspond to distorted spiral, transverse conical spiral, fan-like, canted antiferromagnet, phase I, and forced ferromagnetic, respectively.}
\label{fig:str_phase}
\end{figure}

Shown in Fig.~\ref{fig:str_phase}(a), YMn$_6$Sn$_6$ comprises Mn atoms on a kagome lattice in the $ab$-plane, which are then stacked along the $c$-axis with the layers separated either
by three Sn layers (Sn$_3$) or a mixed Y and Sn layer (Sn$_2$Y). Mn atoms in-plane are at equivalent positions and are strongly coupled ferromagnetically via nearest neighbor exchange ($J_p < 0$) and have the spins in the $ab$-plane due to easy-plane anisotropy ($K < 0$). This stacking pattern has an important magnetic implication, mainly, that within a unit cell there are two unequal interlayer exchange pathways with opposite signs. The interaction across the Sn$_3$ layer is ferromagnetic ($J_1 < 0$) and across the Sn$_2$Y layer, it is antiferromagnetic ($J_2 > 0$). These exchange parameters alone would be compatible with a commensurate antiferromagnetic structure, where the magnetic unit cell is doubled along the $c$-axis. Indeed, this is the initial magnetic structure just below the N\'{e}el temperature ($T_N \approx 340$ K). However, the interlayer coupling between like-Mn layers is ferromagnetic ($J_3 < 0$), and below 333 K the exchange competition drives the system into a double-flat-spiral magnetic structure. In this structure, two rotation angles are needed to describe the directions of the spins (see, for example, Ref.~\onlinecite{ROSENFELD20081898}). One angle defines the relative difference between the two layers of spins within the unit cell, and the second angle defines the relative difference between the layers of spins in adjacent unit cells, and these angles are highly temperature dependent in YMn$_6$Sn$_6$. Curiously, the transition to this incommensurate structure sees two double-flat-spirals emerge: $\mathbf{k}_1 = (0, 0, k_{z,1})$ and $\mathbf{k}_2 = (0, 0, k_{z,2})$, where $k_{z,1}$ and $k_{z,2}$ are almost the same, and both are long-range. \cite{VENTURINI199135, VENTURINI1996102, eichenberger2017, ghimire2020} Single-crystal neutron diffraction data in Figs.~\ref{fig:str_phase}(b) and (c) show the transition from the commensurate to incommensurate structure by tracking the wavevectors and magnetic Bragg peak intensities at the $(1, 0, 0) + \mathbf{k}$ positions. The wavevectors for the incommensurate structures are strongly temperature dependent, getting closer with decreasing temperature, but never merge (at least to 12 K), and they have similar in-field behavior.   

Upon application of an external magnetic field in the $ab$-plane, the magnetic phase diagram becomes much more complex (see Fig.~\ref{fig:str_phase}(d)). A previous study identified five new magnetic phases via ac-susceptibility measurements, \cite{ghimire2020} and through theoretical and neutron diffraction studies was able to predict/confirm the structure of some of those phases. Here we present the solved magnetic structures for two of the in-field phases previously identified but unsolved, namely phase ``II'' -- from here on out denoted canted antiferromagnet (CAF) -- and fan-like (FL), using single-crystal unpolarized neutron diffraction measurements. Additionally, we were able to identify the change in magnetic structure that leads to the region of the ac-susceptibility phase diagram called phase ``I.'' 

We also present a completely new result obtained via a polarized neutron diffraction study. Unexpectedly, unequal chiral domain populations of the zero-field spiral state were found despite the underlying centrosymmetric crystal symmetry. This could be a significant finding as it implies that the spiral state can energetically favor one domain over the other, possibly in a controlled manner. This is another example, along with the THE, of YMn$_6$Sn$_6$ displaying unusual behavior for a structure with inversion symmetry.

\section{Experimental Details}

Single crystals of YMn$_6$Sn$_6$ were grown by the self-flux method described in Ref.~\cite{ghimire2020}, and all neutron experiments used the same 70 mg crystal. For all data, error bars represent plus and minus one standard deviation of uncertainty.

Data for Figs.~\ref{fig:str_phase}(b) and (c) were taken using a single crystal oriented in the $(H,0,L)$ scattering plane on the BT-7 triple-axis spectrometer at the NIST Center for Neutron Research.~\cite{lynn2012double} Elastic diffraction measurements were performed using $E_i=E_f=14.7$ meV with open$-25'-25'-120'$ collimation before the monochromator, sample, analyzer, and detector, respectively. All other neutron data, with the exception of Fig.~\ref{fig:phaseI}, were taken using a single-crystal oriented in the $(H, H, L)$ scattering plane with $25'-25'-25'-25'$ collimation. 

A 10 T superconducting vertical field magnet was used to take in-field measurements where the field was parallel to the crystallographic $[\bar{1},1,0]$ direction. The high sample quality resulted in a sharp mosaic, and data for the magnetic structure determination were taken as $\theta - 2 \theta$ scans through the Bragg peaks. To extract the intensity proportional to the structure factor squared, integrated Bragg peak intensities were corrected by the Lorentz factor ($I_{hkl}^{obs} \propto \frac{\left| F_{hkl} \right| ^2}{\mathrm{sin}2 \theta _{hkl}}$), and these values were used to refine structures with the Rietveld method and the program FullProf.~\cite{FullProf} Measurements of the nuclear Bragg peaks at 0 T revealed that extinction effects, and possibly multiple Bragg scattering, diminished the intensity of the strongest peaks; thus, any magnetic intensity appearing at these positions upon application of the field was excluded from refinement for the in-field structure determinations. 

The beam for the polarized neutron diffraction measurements was created using a $^{3}$He polarizer before the sample, and polarization analysis was made possible using an additional $^{3}$He polarizer after the sample.~\cite{CHEN2007168} A guide field of 1 mT was employed to define the polarization axis and was oriented in-plane and along the scattered wave vector, or perpendicular to the scattering plane. Initial flipping ratios were typically $\approx 34$. The four neutron scattering cross-sections available for measurement were $I^{++}$, $I^{+-}$, $I^{-+}$, and $I^{--}$.  Data were taken with the scattering vector, $\mathbf{Q}$, both parallel and perpendicular to the neutron polarization, $\mathbf{P}$, and the temperature was held constant at 290 K. All data were corrected for polarization efficiency before analysis. Both nuclear and magnetic Bragg peaks were resolution limited, and Voigt functions were used to fit the data. 

\begin{figure}[b]
\includegraphics[scale=0.155]{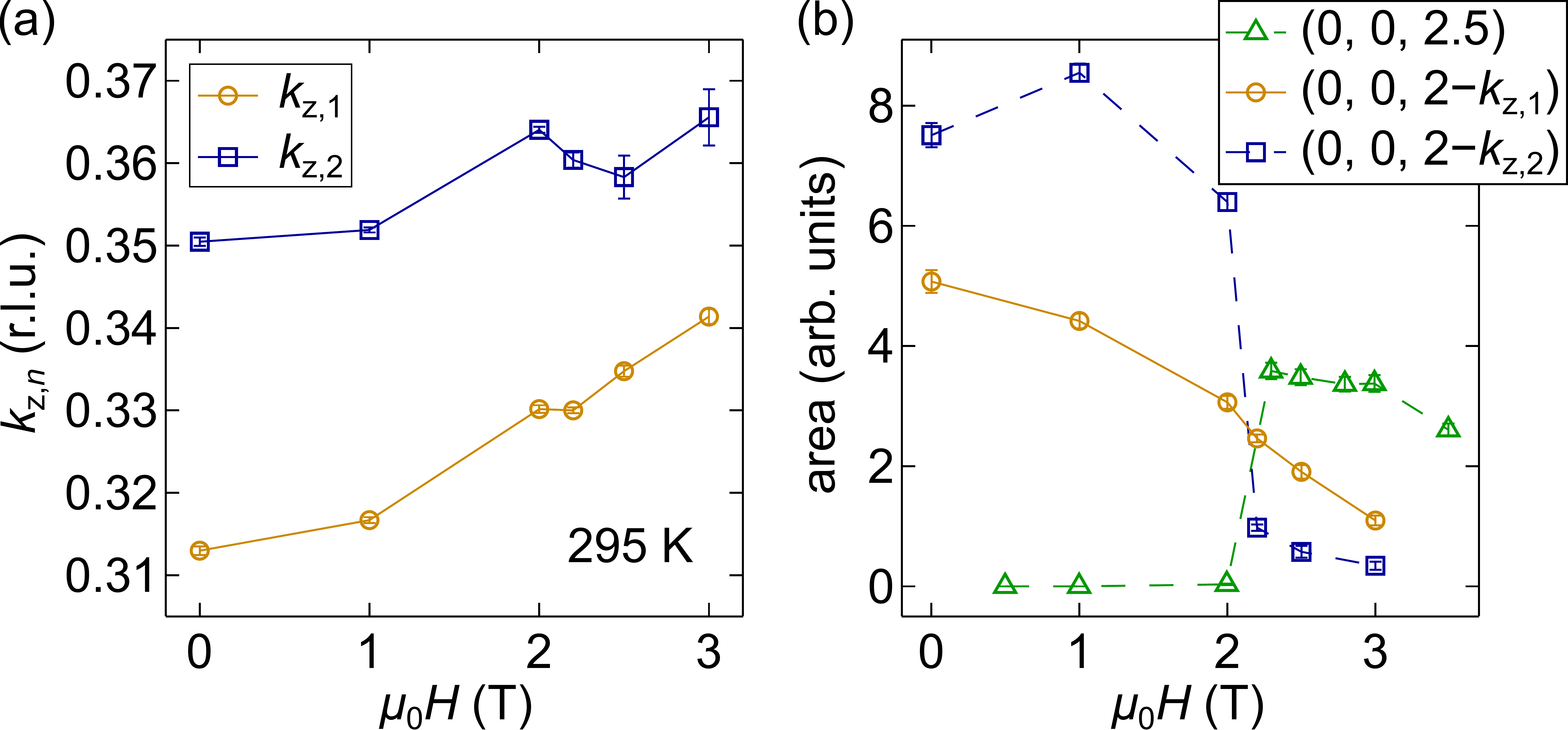}
\caption{Room temperature data showing the evolution of the magnetic structures with increasing applied field. (a) The change in periodicity for the two incommensurate structures with wavevectors, $(0, 0, k_{z,n})$. (b) The change in intensity for magnetic Bragg peaks about the $(0, 0, 2)$ reciprocal lattice point. As the Bragg peak at $(0, 0, 2-k_{z,2})$ rapidly decreases in intensity above 2 T, the commensurate Bragg peak at $(0, 0, 2.5)$ just as rapidly increases in intensity. The dashed lines for both data sets are to emphasize the relationship between the two, which suggests the incommensurate $k_{z,2}$ structure is transitioning into the commensurate structure above 2 T.}
\label{fig:RTLFOP}
\end{figure}

High resolution synchrotron powder diffraction data were collected using beamline 11-BM at the Advanced Photon Source at Argonne National Laboratory using a wavelength of 0.4579 {\AA}. Due to the high absorption of Sn at this wavelength, samples were prepared by coating the outside of a 0.8 mm diameter Kapton capillary with a mixture of sample powder (a ground single crystal of YMn$_6$Sn$_6$) and Dow Corning 4 Electrical Insulating Compound silicone grease. Refinement of the data was performed using the program FullProf.~\cite{FullProf} All data sets (temperatures) were first refined using a Lebail fit in order to obtain the lattice and peak profile parameters and the background. It was found that the peaks could be fully described by a Lorentzian profile and that some peak width anisotropy was present, where ($0,0,L$)-type peaks tended to be slightly narrower than others. A spherical harmonics size-broadening model was able to capture the peak profile shape correctly for all peaks. The profile and background parameters were then used, and held constant, for the Rietveld refinement. Lattice parameters, anisotropic atomic displacement parameters, and Sn occupancies were allowed to refine. An impurity phase from elemental Sn, which was used during flux growth, was also included in the refinement, and found to be $\approx 7 \%$. 

\section{Results}

\subsection{In-field magnetic structures}

\subsubsection{Room temperature, low-field canted antiferromagnetic phase}

Previous ac susceptibility and neutron diffraction measurements identified a small region of finite field-temperature phase space with a commensurate magnetic structure and wavevector of $(0, 0, 0.5)$. \cite{ghimire2020} The phase was labeled ``II'' and was stabilized at fields ranging between $\approx 2$ T and 4 T and spanned temperatures between $\approx 250$ K and 320 K. We have studied the field-dependent onset of the phase at 295 K and have solved the magnetic structure at 3 T.  

An important note is that all incommensurate phases are present with two wavevectors, $\mathbf{k}_1 = (0, 0, k_{z,1})$ and $\mathbf{k}_2 = (0, 0, k_{z,2})$, where $\left| k_{z,1} \right| < \left| k_{z,2} \right|$ for all temperatures with and without applied magnetic field. Due to the proximity to each other, the high resolution measurements presented here are needed to resolve the Bragg peaks associated with each wavevector. As such, the periodicity of the incommensurate wavevectors was tracked as a function of field, shown in Fig.~\ref{fig:RTLFOP}(a). For both incommensurate structures, the period of the spiral is generally shortened with increasing field, with the exception of a short-lived increase between 2 T and 2.2 T. It is between these fields that the commensurate structure abruptly emerges, as shown by the field-dependent intensity data of the $(0, 0, 2.5)$ magnetic Bragg peak in Fig.~\ref{fig:RTLFOP}(b). As the commensurate structure sets in, the incommensurate structure associated with $k_{z,2}$ loses most of its intensity, indicating a phase transition of this incommensurate structure to the commensurate one. Meanwhile, the incommensurate structure associated with $k_{z,1}$ monotonically and smoothly decreases in intensity with applied field. 

\begin{figure}[t]
\includegraphics[scale=0.2]{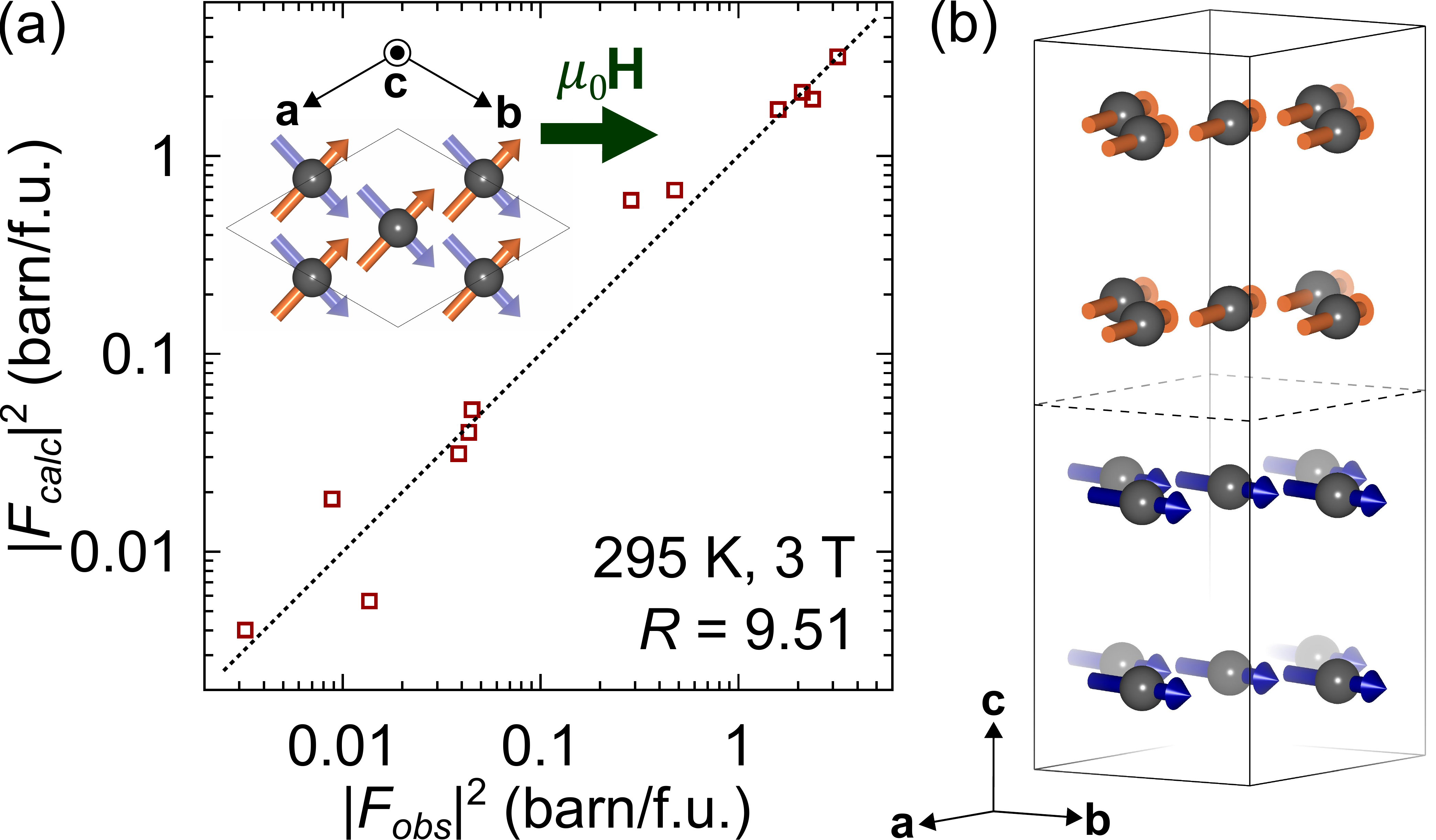}
\caption{Results of the best fit magnetic model for the room temperature, low-field canted antiferromagnetic phase with wavevectors, $(0, 0, 0)$ and $(0, 0, 0.5)$. Data were taken at 295 K and 2 T in the $(H, H, L)$ scattering plane, and the figures represent the magnetic structure at this particular field and temperature. (a) The observed versus calculated magnetic structure factor squared. The calculated result was obtained via Rietveld refinement for the magnetic intensity only. The inset shows the magnetic structure in the $ab$-plane. Only Mn ions are shown (grey spheres), and the orange and blue arrows represent the two different directions of the moments. (b) The magnetic unit cell, where the dashed line defines the size of nuclear unit cell.}
\label{fig:RTLFModel1}
\end{figure}

The Rietveld refined fit and structure are depicted in Fig.~\ref{fig:RTLFModel1}. The best-fit magnetic structure was found to have the same AF coupling as the high-temperature, zero-field structure which initially sets in with the onset of long-range order at $T_N$. That is, magnetic ions through the Sn$_3$ layer are FM coupled, and ions through the Sn$_2$Y layer are AFM coupled. All ions within a layer are FM coupled, as is the case for all the reported YMn$_6$Sn$_6$ magnetic phases. Due to the applied field, the moments are all canted towards the field direction, adding a net ferromagnetic component and second commensurate wavevector, $\mathbf{k}=0$. The angle the moments make with the applied field direction is denoted $\gamma$. The moments through the Sn$_2$Y layer (the AFM coupled layers) were constrained during refinement such that the angles away from $\mathbf{H}$ both had a magnitude of $\gamma$ and all moments were constrained to have the same magnitude. The refined angle, $\gamma = 52^{\circ} \pm 2^{\circ}$, and the refined moment, $\mu = 1.13 \mu _{\mathrm{B}} \pm 0.04 \mu _{\mathrm{B}}$, resulted in a fit with an $R$-factor of $9.21$. 

\subsubsection{Low temperature, high-field fan-like phase}

The low-temperature, high-field commensurate magnetic phase was denoted ``fan-like'' (FL) in Ref.~\onlinecite{ghimire2020}. It can be described with wavevectors $(0, 0, 0)$ and $(0, 0, 0.25)$, and an additional modulation within the $4c$ periodicity resulted in $(0, 0, 0.5)$-type magnetic Bragg peaks. The region of phase space spanned by this phase is much larger than the CAF phase previously discussed. Data presented here were taken at 1.5 K and 7.8 T, where there was no trace of any incommensurate structure. 

\begin{figure}[b]
\includegraphics[scale=0.23]{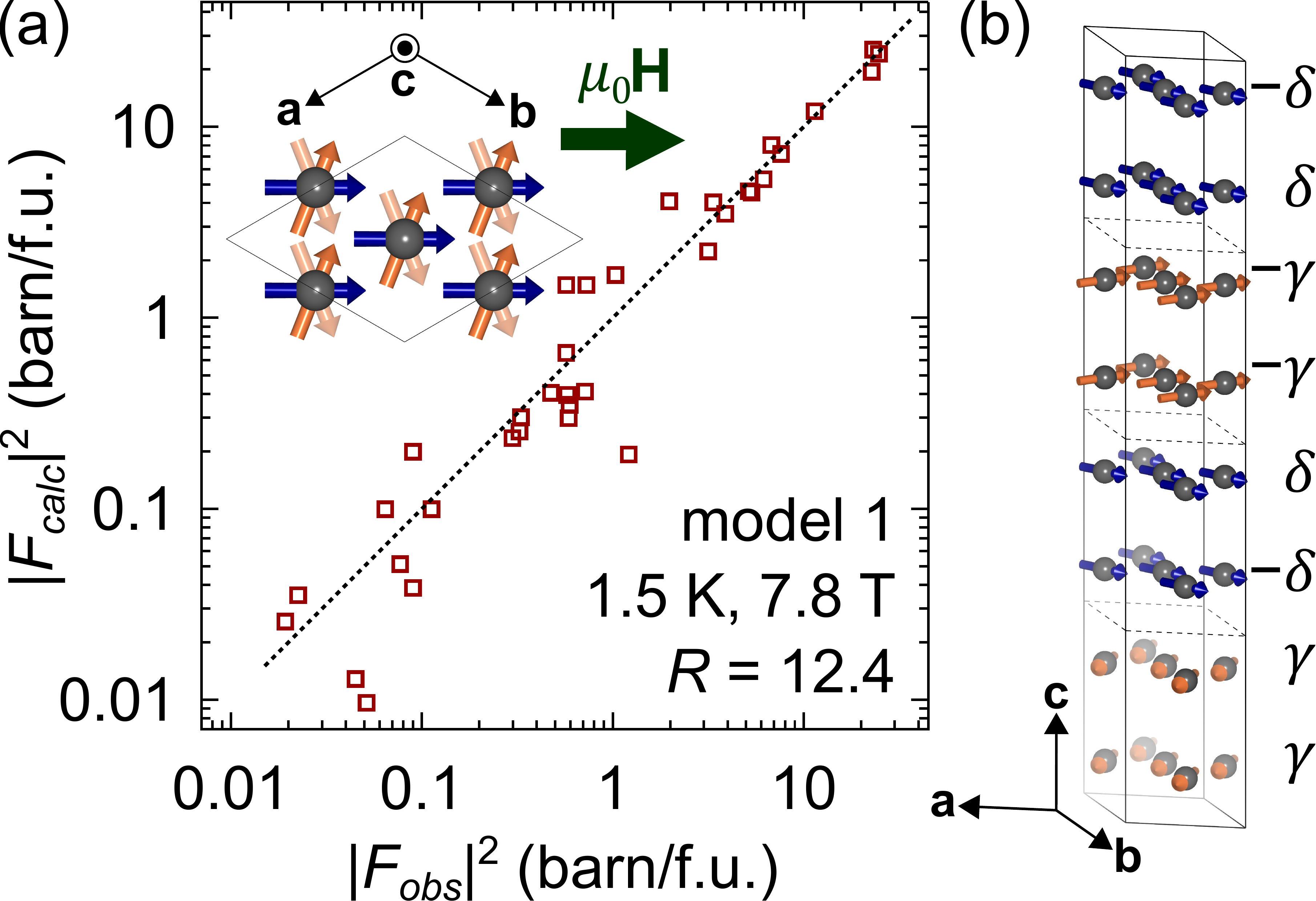}
\caption{Results for ``model 1'' of the low temperature, high-field fan-like phase with wavevectors, $(0, 0, 0)$, $(0, 0, 0.25)$, and $(0, 0, 0.5)$. Data were taken at 1.5 K and 7.8 T in the $(H, H, L)$ scattering plane. (a) The observed versus calculated magnetic structure factor squared. The calculated result was obtained via Rietveld refinement for the magnetic intensity only. The inset shows the refined magnetic structure in the $ab$-plane for 1.5 K and 7.8 T, where $\gamma = 68^{\circ} \pm 2^{\circ}$ and $\delta = 0^{\circ} \pm 1^{\circ}$. Only Mn ions are shown (grey spheres), and the orange and blue arrows represent the two different angle magnitudes, $\gamma$ and $\delta$, respectively, which define the moment directions away from the applied field direction. (b) The magnetic unit cell, where the dashed lines define the size of the nuclear unit cell.}
\label{fig:LTHFModel1}
\end{figure}

The theoretical model in Ref.~\onlinecite{ghimire2020} found a stable magnetic structure matching the periodicity of the observed magnetic Bragg peaks. The moment directions for the eight layers of Mn atoms within the magnetic unit cell could be described by angles, $\gamma$, $\gamma$, $- \delta$, $\delta$, $- \gamma$, $- \gamma$, $\delta$, $- \delta$, which are measured with respect to the field direction. This structure is viewed in Fig.~\ref{fig:LTHFModel1} where orange and blue arrows represent moments whose directions can be defined by either the angle $\gamma$ or $\delta$, respectively. The refined fit for this model (``model 1'') is shown in Fig.~\ref{fig:LTHFModel1}(a) with $\gamma = 68^{\circ} \pm 2^{\circ}$, $\delta = 0^{\circ} \pm 1^{\circ}$, $\mu = 1.95 \mu _{\mathrm{B}} \pm 0.05 \mu _{\mathrm{B}}$, and an $R$-factor of 12.4. All moments were constrained to have the same magnitude. 

Another model, ``model 2'', resulted in a similar goodness-of-fit ($R$-factor = 12.1, see Appendix). The relationship between angles in this model can be described as $\gamma$, $\gamma$, $- \delta$, $- \delta$, $- \gamma$, $- \gamma$, $\delta$, $\delta$, with refined values $\gamma = 69^{\circ} \pm 3^{\circ}$, $\delta = 12^{\circ} \pm 6^{\circ}$, and $\mu = 1.91 \mu _{\mathrm{B}} \pm 0.06 \mu _{\mathrm{B}}$. The magnetic structure factors for both models are almost identical. If $\gamma$, $\delta$, and $\mu$ were the same for both models, the structure factors for $(0, 0, 0)$-type and $(0, 0, 0.5)$-type peaks would also be the same. The structure factors would only differ for the $(0, 0, 0.25)$-type peaks, but as $\delta \rightarrow 0$ in model 1, the structure factors for these peaks converge to that of model 2. Details of the structure factor calculations can be found in the Appendix. 

\subsubsection{Phase I}
\begin{figure}[t]
\includegraphics[scale=0.24]{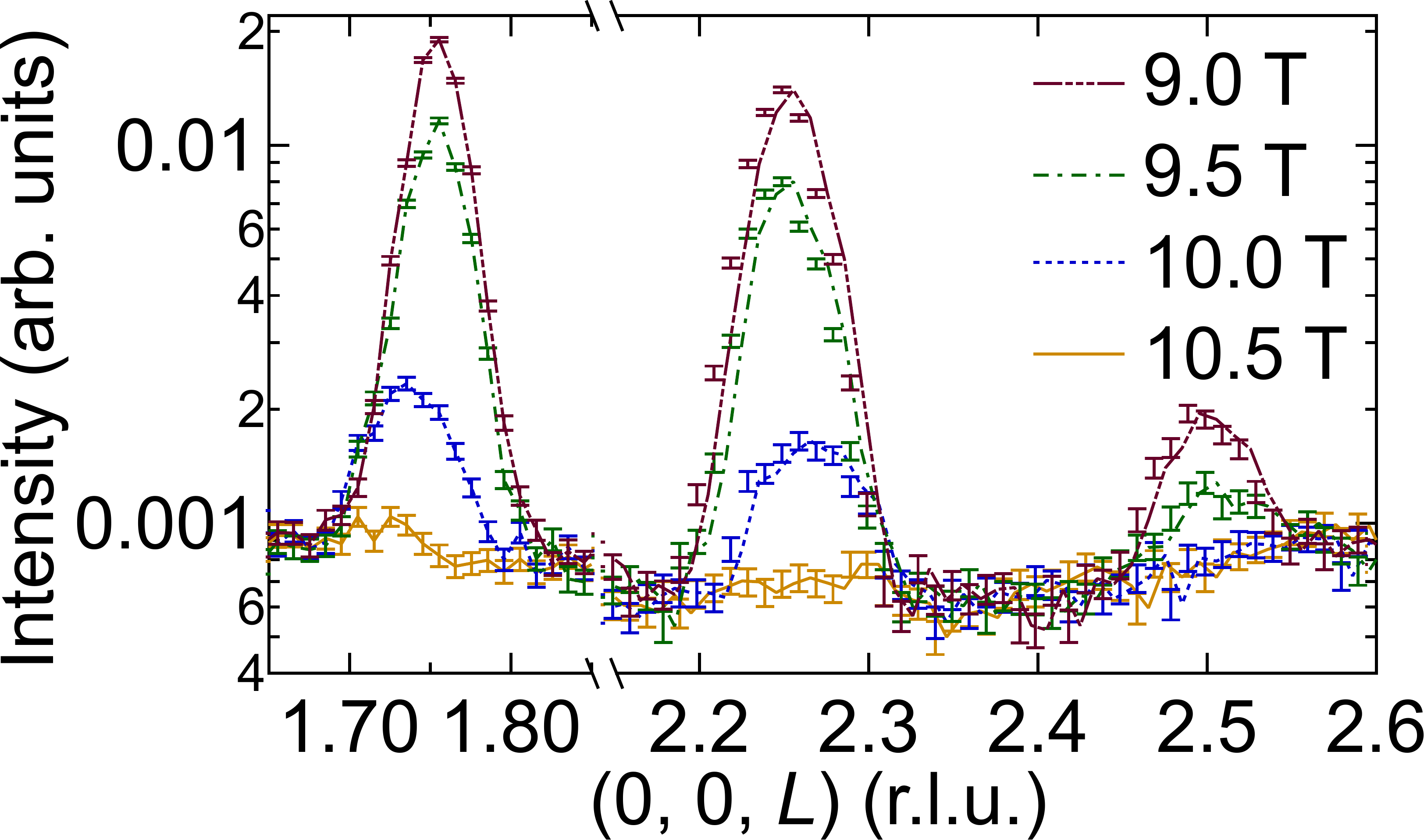}
\caption{High-field data with $\mathbf{H} \parallel [\bar{1}, 1, 0]$, taken at 10 K with a position sensitive detector and moderately course resolution. Between the FL phase (9.0 T and 9.5 T) and FF phase (10.5 T), an intermediate phase appears, where the 0.5-like peaks disappear, and the magnetic structure is incommensurate.}
\label{fig:phaseI}
\end{figure}

We now comment on the ``I'' region of the phase diagram in Fig~\ref{fig:str_phase}(b). Data in Fig.~\ref{fig:phaseI} were taken at 10 K and show an intermediate magnetic structure between the FL phase, at 9.0 T and 9.5 T, and the FF phase, at 10.5 T. At 10.0 T, the 0.5-type Bragg peak at (0, 0, 2.5) is completely gone, and the peaks at 0.25-type positions have shifted away from the zone centers to become, once again, incommensurate. The positions of the incommensurate peaks shown are at $L = 1.7335 \pm 0.0008$ and $2.266 \pm 0.001$, which correspond to an average wavevector of $\mathbf{k} \approx (0, 0, 0.26)$, the same as the average of the two zero-field wavevectors at this temperature. \cite{ghimire2020} These data were taken with a position sensitive detector and moderately course resolution (open$-50'-40'R-120'$, where ``\textit{R}'' indicates radial), where any splitting of the peaks would not be resolvable. 

\subsection{Synchrotron powder diffraction}

\begin{figure}[t]
\includegraphics[scale=0.28]{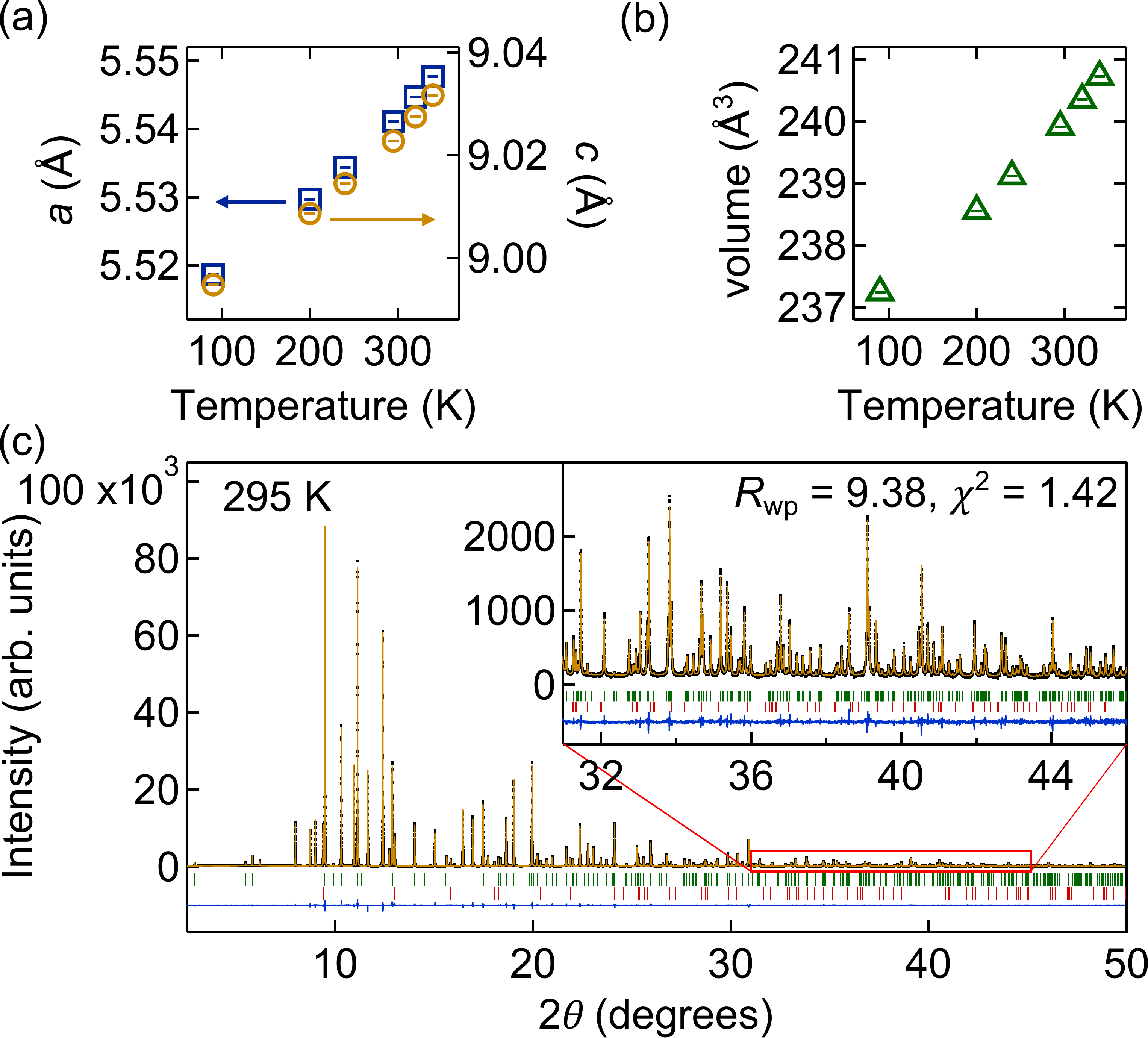}
\caption{Results from the analyzed synchrotron powder diffraction data. (a) The lattice parameters and (b) unit cell volume versus temperature. (c) Rietveld refinement for 295 K data. The data are displayed as black dots and the Rietveld calculated fit is the solid yellow line running through the data. The top row of tik marks (green) denote YMn$_6$Sn$_6$ Bragg peak positions, and the bottom row of tik marks (red) are the elemental Sn impurity Bragg peak positions. The difference curve (observed$-$calculated) is shown as the solid blue line on the bottom of the plot. The inset highlights the fit in the high-Q region. }
\label{fig:xray}
\end{figure}

One of the intriguing magnetic properties of YMn$_6$Sn$_6$ is the observation of two distinct incommensurate wave vectors for the zero-field magnetic structure, which has been observed in essentially all the magnetic neutron studies. \cite{VENTURINI199135, VENTURINI1996102, eichenberger2017, ghimire2020} One obvious explanation would be that the samples grow in two slightly different structures or compositions, so there are two different samples under investigation. To ascertain if this might be the case, we carried out high resolution synchrotron powder diffraction measurements to determine if more than one set of lattice parameters coexists, which could explain the presence of the two slightly different incommensurate magnetic modulations. However, the results definitively show that only one set of lattice parameters explains the data, and these are shown in Fig.~\ref{fig:xray}(a). Excellent fits to the data were obtained, and we find that the lattice parameters and volume (Fig.~\ref{fig:xray}(b)) monotonically decrease with decreasing temperature, showing no discontinuity. Fig.~\ref{fig:xray}(c) shows an example of the calculated Rietveld refinement and data for 295 K. The refined parameters for all the data sets are displayed in Table \ref{tab:xray}. There may be evidence for some slight inhomogeneity in the Sn content, since two data sets, at temperatures 90 K and 295 K, were taken at a slightly different conditions and different sampling positions than the rest. The total refined Sn content was found to be slightly lower for these two temperatures, with the main difference being the occupancy at the Sn3 site.

\begin{table*}[t]
\begin{tabular}{@{}lllllllllll@{}}
\toprule
\textbf{\textit{T} (K)} & \textbf{\textit{R}$_{wp}$} & $\mathbf{\chi ^2}$ & \textit{\textbf{a ({\AA})}} & \textit{\textbf{c ({\AA})}} & \textbf{Mn \textit{z} (\textit{z}/\textit{c})} & \textbf{Sn3 \textit{z} (\textit{z}/\textit{c})} & \textbf{Sn1 occ.} & \textbf{Sn2 occ.} & \textbf{Sn3 occ.} & \textbf{YMn}$_6$\textbf{Sn}$_x$ \\ \midrule
90$^*$    & 9.40 & 1.70   & 5.518671(1)             & 8.994806(2)             & 0.24753(4)          & 0.33689(3)           & 97.12(8)                    & 96.86(8)                    & 97.83(9)                    & 5.836(3)         \\
200       & 10.0 & 2.24   & 5.529688(1)             & 9.008645(2)             & 0.24750(4)          & 0.33700(3)           & 98.16(9)                    & 98.20(9)                    & 99.68(9)                    & 5.921(3)         \\
240       & 9.93 & 2.12   & 5.534370(2)             & 9.014464(3)             & 0.24739(4)          & 0.33695(3)           & 98.30(8)                    & 98.26(9)                    & 99.64(9)                    & 5.924(3)         \\
295$^*$   & 9.38 & 1.42   & 5.5410810(9)            & 9.022760(2)             & 0.24719(4)          & 0.33724(3)           & 97.74(8)                    & 97.27(8)                    & 98.32(8)                    & 5.867(3)         \\
320       & 9.98 & 1.94   & 5.544690(1)             & 9.027480(2)             & 0.24730(4)          & 0.33729(3)           & 98.05(8)                    & 97.99(9)                    & 99.94(9)                    & 5.919(3)         \\
340       & 9.60 & 1.78   & 5.547689(1)             & 9.031668(2)             & 0.24755(4)          & 0.33709(3)           & 97.88(8)                    & 97.97(9)                    & 100.08(9)                   & 5.919(3) \\ \bottomrule
\end{tabular}
\caption{\label{tab:xray}Rietveld refined parameters from synchrotron powder diffraction data. The space group used for refinement was $P6/mmm$ (191) and the atomic positions are: Y (0, 0, 0), Mn ($\frac{1}{2}$, 0, $z$), Sn1 ($\frac{1}{3}$, $\frac{2}{3}$, $\frac{1}{2}$), Sn2 ($\frac{1}{3}$, $\frac{2}{3}$, 0), and Sn3 (0, 0, $z$). Temperatures denoted with $^*$ were taken at a slightly different sampling position than the rest of the temperatures.}
\end{table*}

\begin{figure*}
\includegraphics[scale=0.16]{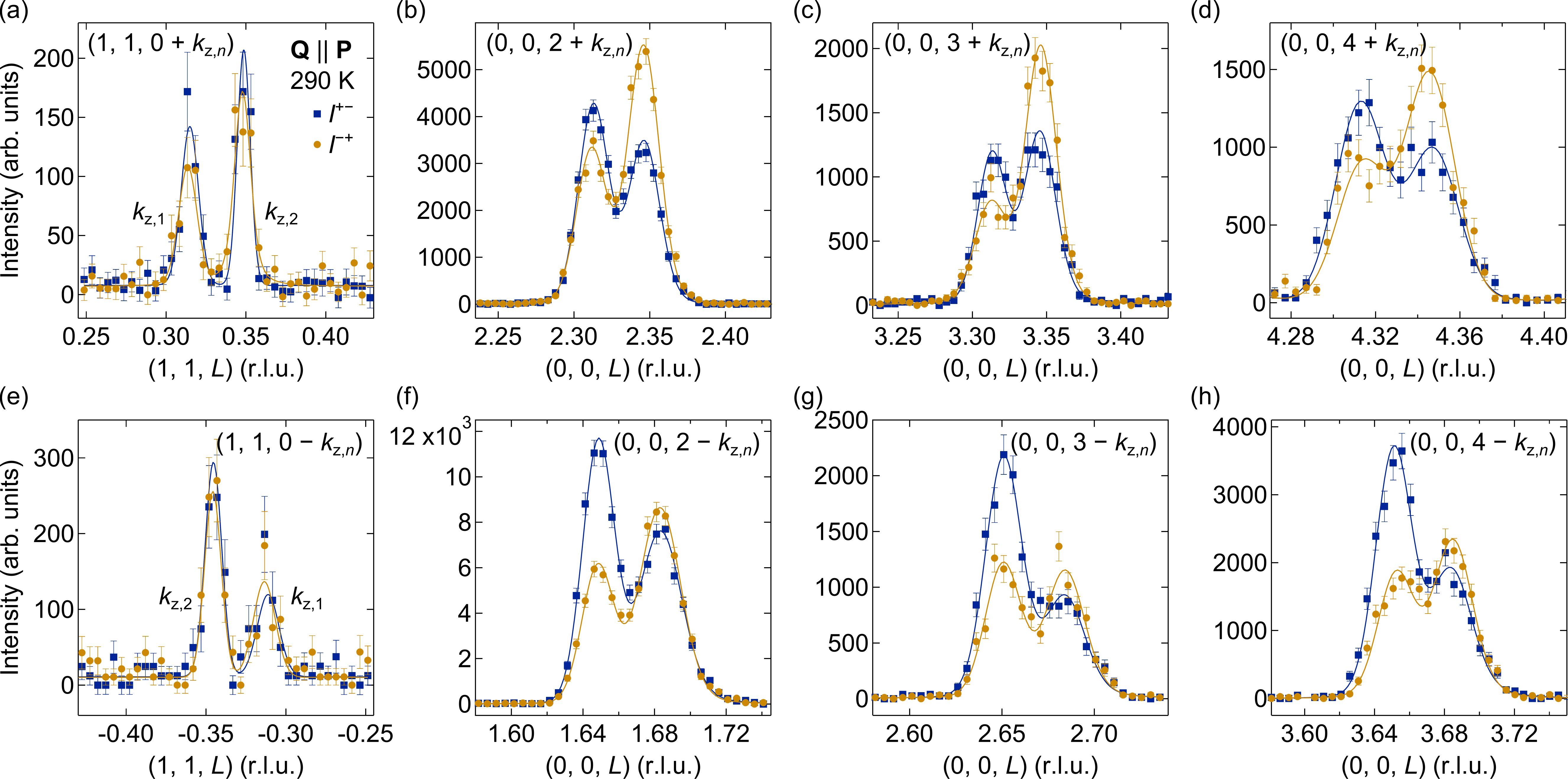}
\caption{Polarized neutron data with $\mathbf{Q} \parallel \mathbf{P}$ at $T = 290$ K showing scans which span the magnetic Bragg peaks associated with the two wavevectors, $\mathbf{k}_i$ ($i = 1, 2$,  $\left| \mathbf{k}_1 \right| < \left| \mathbf{k}_2 \right|$), discussed in the main text. Each panel displays the two spin-flip channels measured: $+-$ (blue squares) and $-+$ (orange circles). The top row, (a)-(d), shows Bragg peaks with zone-center $+\mathbf{k}_i$ momenta, and the bottom row, (e)-(h), shows the corresponding zone-center $-\mathbf{k}_i$ momenta. }
\label{fig:polneut1}
\end{figure*}

\subsection{Polarized neutron analysis at 290 K and 0 T}
\subsubsection{$\mathbf{Q} \parallel \mathbf{P}$}

\begin{figure}[t]
\includegraphics[scale=0.175]{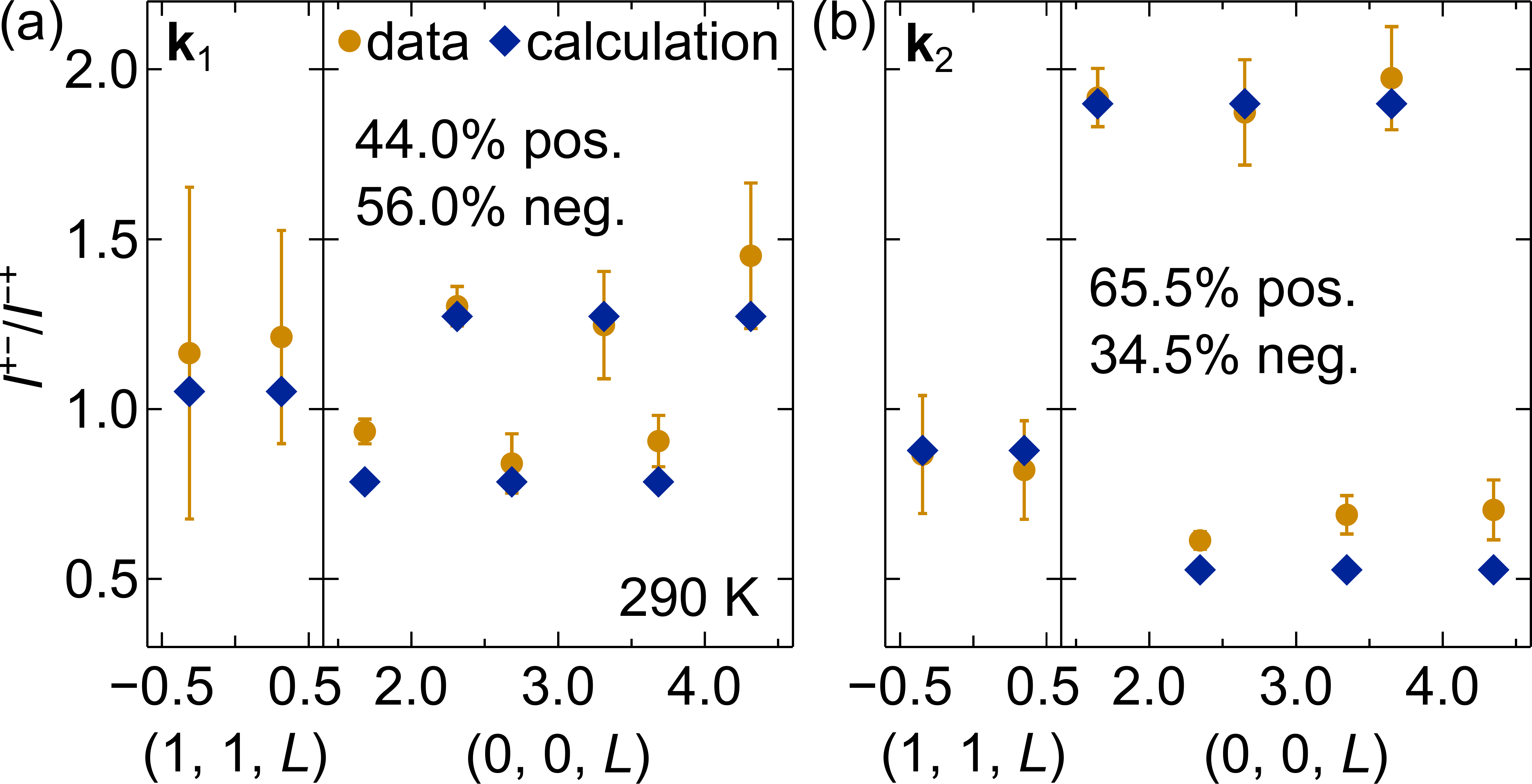}
\caption{Ratios of the $\mathbf{Q} \parallel \mathbf{P}$ polarized neutron spin-flip channels, $I^{+-}/I^{-+}$, for the (a) $\mathbf{k}_1$ wavevector and (b) $\mathbf{k}_2$ wavevector. Ratios from the data, taken at $T = 290$ K and shown as orange circles, represent the integrated intensities of fits to the data, as described in the main text. The calculated ratios, shown as blue diamonds, come from structure factor calculations for the given chiral domain populations.}
\label{fig:polneut2}
\end{figure}

The spin-flip (SF) cross-section intensities, $I^{+-}$ and $I^{-+}$, for magnetic Bragg peaks stemming from multiple zone centers are shown in Fig.~\ref{fig:polneut1}, with the fits to the data shown as solid lines. The non-spin-flip (NSF) cross-section intensities, $I^{++}$ and $I^{--}$, were also measured, but yielded no intensity, as expected for magnetic Bragg peaks in the $\mathbf{Q} \cdot \mathbf{P} = 1$ configuration defined by a guide field of 1 mT. For each panel in Fig.~\ref{fig:polneut1}, both wavevectors, $\mathbf{k}_1$ and $\mathbf{k}_2$ are covered via a scan along the $L$-direction, revealing that for a given wavevector, the SF cross-section that is most intense appears to depend on whether the peak is on the higher- or lower-Q side of a given zone-center. Scans along the $HH$-direction in this $(H,H,L)$ scattering plane were also performed to ensure the peaks were centered at the commensurate position in that direction. The integrated areas for each wavevector and cross-section were evaluated and the ratios, $I^{+-}/I^{-+}$, are plotted for $\mathbf{k}_1$ and $\mathbf{k}_2$ in Figs.~\ref{fig:polneut2}(a) and (b), respectively. As discussed further in the Discussion section, the only way for the two SF cross-section intensities to differ is in the presence of a spiral-type structure. Typically, in a centrosymmetric crystal, one wouldn't see this difference due to multiple magnetic domains being evenly populated, and the difference here is due to the uneven population of the two possible chiral domains, referred to here as positive or negative chirality, where the chiral sign is defined by the sign of $\mathbf{S}_i \times \mathbf{S}_j$ (where $i$ and $j$ here refer to nearest neighbor non-collinear spins along the propagation direction). A least-squares calculation was performed to find the percentage of each chiral domain which best fit the $I^{+-}/I^{-+}$ ratio data, and the results are shown as blue diamonds in Fig.~\ref{fig:polneut2}. The dominant chirality for $\mathbf{k}_1$ was found to be negative at 56.0\%, and the dominant chirality for $\mathbf{k}_2$ was found to be positive at 65.5\%. Note that the chiralities for the two spirals are opposite, and comparable in magnitude, meaning that one spiral propagated (preferentially) in one direction, and the other in the opposite direction. 

\subsubsection{$\mathbf{Q} \perp \mathbf{P}$}

\begin{figure}[t]
\includegraphics[scale=0.15]{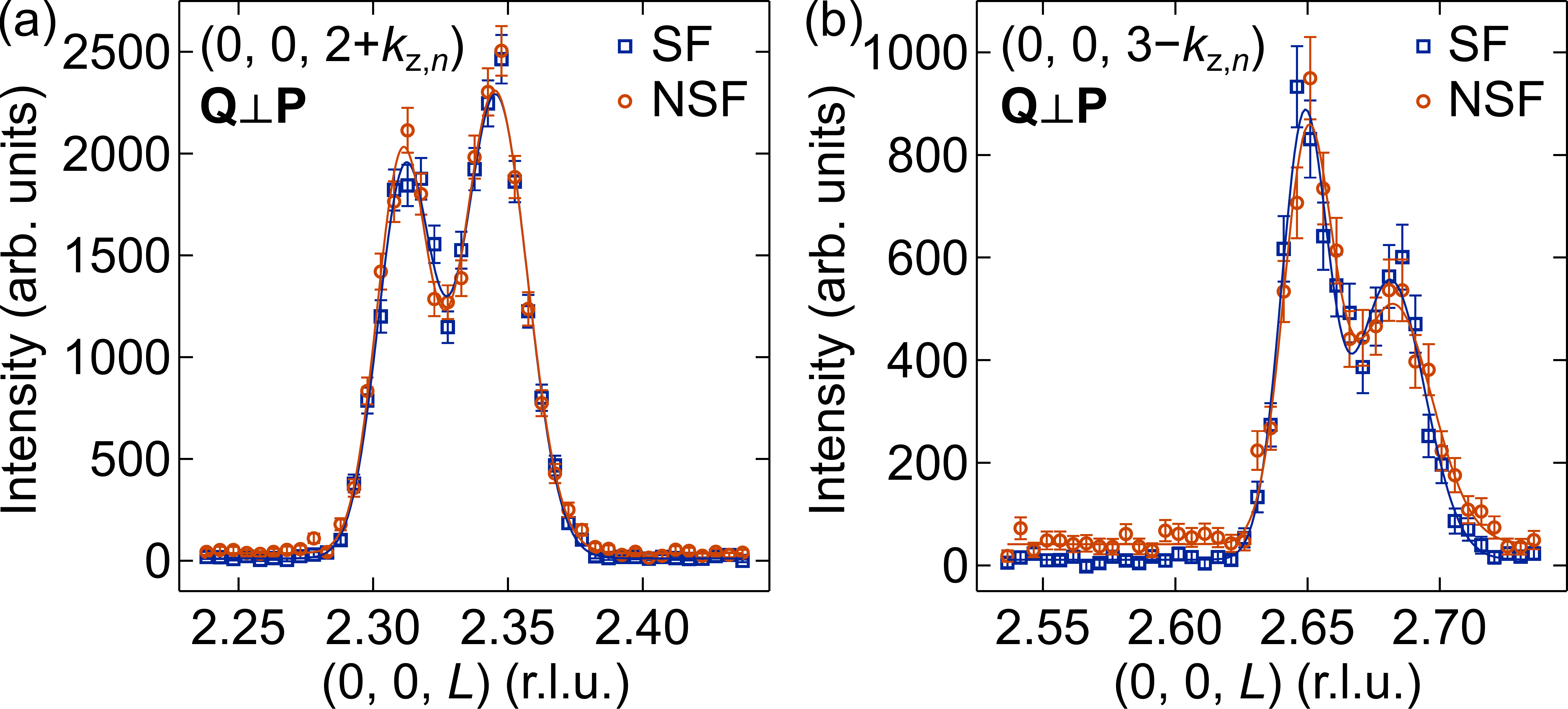}
\caption{Polarized neutron data with $\mathbf{Q} \perp \mathbf{P}$ at 290 K for (a) $(0, 0, 2+k_{z,n})$ and (b) $(0, 0, 3-k_{z,n})$. For both (a) and (b), there is no difference between the integrated intensity of the SF and NSF data. This implies that the moments trace a circle as they spiral along the $c$-axis.}
\label{fig:QperpP}
\end{figure}

In addition to the polarized neutron experiment configuration with the neutron polarization parallel to the scattering vector ($\mathbf{Q} \parallel \mathbf{P}$), we also took data with the polarization perpendicular to the scattering vector and scattering plane ($\mathbf{Q} \perp \mathbf{P}$). There is no chiral term in any of the scattering cross-sections for this configuration, but magnetic scattering is allowed in the non-spin flip (NSF) channel when there is a component of the spin parallel to the polarization vector; this gives directional information about the spin. Fig.~\ref{fig:QperpP} shows the results of the data taken in this polarized geometry at 290 K for the magnetic Bragg peaks, (a) $(0,0,2+k_{z,n})$ and (b) $(0,0,3-k_{z,n})$. There was no difference in intensity between the data from the two spin-flip (SF) channels ($++$ and $--$) or between the two NSF channels ($+-$ and $-+$) for this polarization geometry, and thus the data from the two SF channels were averaged as well as the data from the two NSF channels. For both (a) and (b), there is also no difference between the integrated intensity of the SF and NSF data. This implies that the moments trace a circle as they spiral along the $c$-axis, as opposed to an ellipse. 

\section{Discussion}
Currently, there is not a satisfactory explanation for the co-existence of the two, almost equivalent, wavevectors found in YMn$_6$Sn$_6$ and in some doped variants. \cite{VENTURINI199135, VENTURINI1996102, eichenberger2017, ghimire2020} One possibility suggested was that the magnetic structure has a non-constant rotation of the moments, and the wavevectors observed were merely harmonics of a much smaller fundamental wavevector. \cite{VENTURINI1996102} However, recent inelastic neutron scattering measurements show that the observed wavevectors are, in fact, the magnetic zone center, \cite{zhang2020topological} making the modulated structure theory obsolete. An inhomogenous distribution of two magnetic structures, which are almost energetically identical, could also be likely. Multiple ground states have been observed in intermetallics due to off-stoichiometry, such as the fluctuating Ni concentration in CeNi$_{0.84}$Sn$_2$ which leads to two co-existing magnetic ground states, \cite{Schobinger_Papamantellos_1996} or the ground state sensitivity to the Sn content in Ce$_3$Rh$_4$Sn$_{13}$. \cite{Sfilebarski_Ce3Rh4Sn13} Most similar is the itinerant antiferromagnet, Mn$_3$Sn, where two helical modulations coexist over a wide temperature range. \cite{cable1993neutron} The incommensurate transition temperature and wavevector values were also shown to have a dependence on the annealing history, implying disorder may play a role in the magnetic structure. \cite{marcus2018thesis} However, our synchrotron powder diffraction data show that if chemical inhomogeneity were the root cause for the double wavevectors, then there is no associated structural inhomogeneity in the form of a distribution of lattice parameters, and our neutron diffraction data show that the regions of homogeneous chemical compositions would have to be large enough to lead to long-range magnetic order (i.e.\ $>1000$ {\AA}). 

Also of note is the temperature dependence of the spirals' modulation lengths; the percent change in wavevector component $k_{z,n}$ between the onset of the incommensurate phase at 333 K and the base temperature measured (12 K in Ref.~\cite{ghimire2020}) is quite large: $-29$ \% for $n=1$ and $-43$ \% for $n=2$. Likely, this is due to the sensitivity of the spiral structures to the relative exchange pathway strengths $J_1 - J_3$, which in turn are temperature dependent due to the known importance of thermal fluctuations in this system.

The previous mapping of $k_{z,1}$ and $k_{z,2}$ with applied field in the $ab$-plane \cite{ghimire2020} demonstrates that both magnetic structures are very close in energy to one another, with $k_{z,1}$ consistently undergoing transitions at a slightly lower field than $k_{z,2}$. The room temperature, low-field structure is an exception. The rapid disappearance of the $k_{z,2}$ structure at 2 T as the commensurate structure appears suggests that $k_{z,2}$ is transitioning to the commensurate structure, while the incommensurate $k_{z,1}$ structure smoothly transitions to the forced ferromagnetic (FF) state. The in-field commensurate structure is very similar to that at the N\'{e}el temperature, where the magnetic layers within a unit cell (across the Sn$_3$ layer) are ferromagnetically coupled, and across the Sn$_2$Y layer are antiferromagnetically coupled. The in-field structure reported here is a canted variation of that structure, where all moments simply contribute to a net ferromagnetism pointed in the direction of the applied field.

Another deviation from the lower temperature behavior is the absence of a spin-flop transition as field increases at 295 K. An antiferromagnet with magnetocrystalline anisotropy will have a spin-flop transition at a field proportional to $\sqrt{ \left< J \right> K }$, where in YMn$_6$Sn$_6$, $\left< J \right>$ is the average out-of-plane Heisenberg exchange and $K$ is the easy-plane magnetocrystalline anisotropy. It's expected that as temperature increases the spin-flop field would decrease, as observed via ac susceptibility measurements, but instead, the distorted spiral to commensurate canted antiferromagnet shows no sign of a spin-flop transition, marked by the absence of any $c$-axis component in the reported structure. One explanation could lie in results from a small angle neutron scattering study, which observed clear quasi-2D behavior in the form of spatial ferromagnetic fluctuations in the same temperature range as the absence of spin-flop behavior. \cite{bykov2015quasi} The report concludes that between $\approx 260$ K and the paramagnetic state, the system can best be described as a quasi-two-dimensional fluctuating ferromagnet, despite the spin structure having a net zero moment. 

\begin{figure}[t]
\includegraphics[scale=0.25]{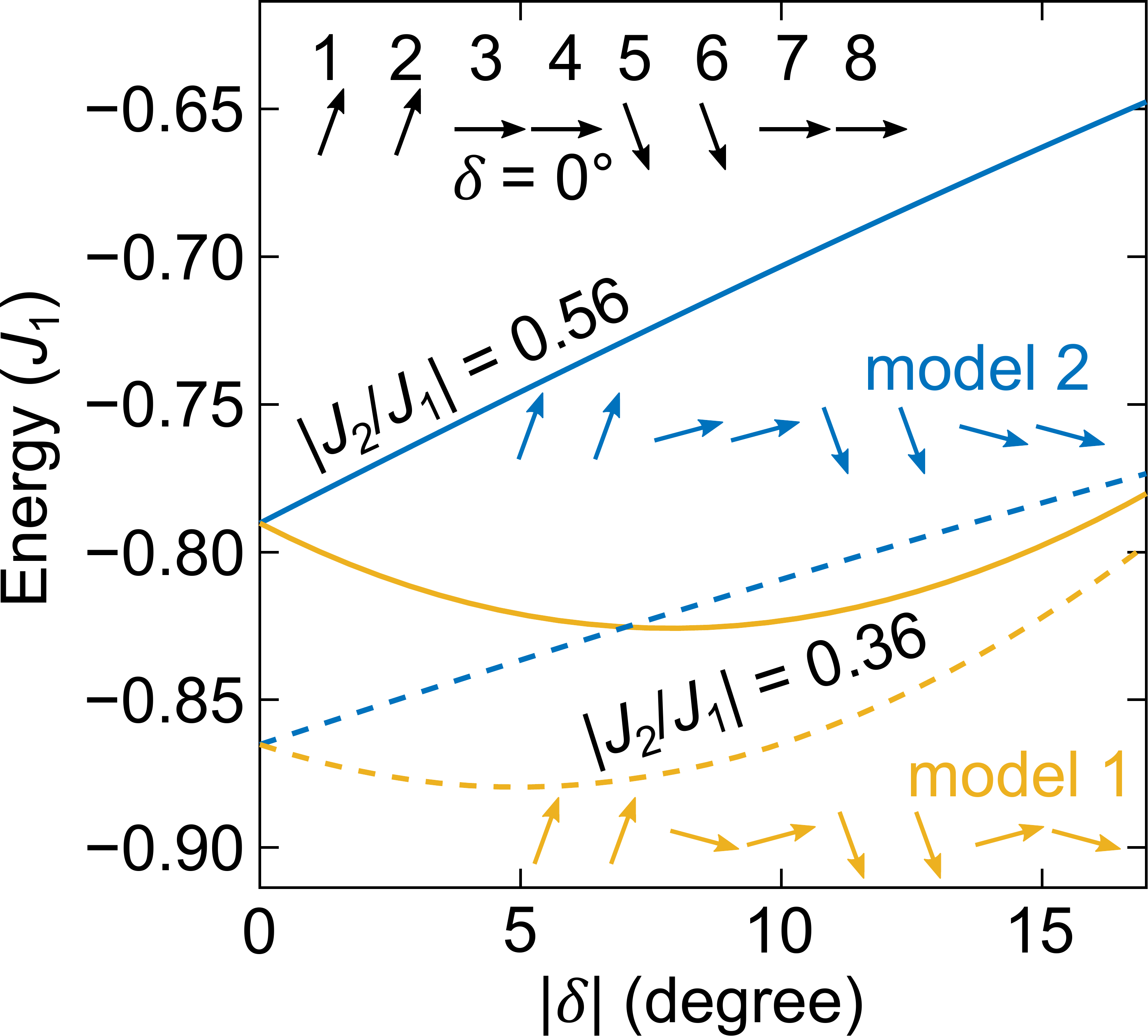}
\caption{Energy comparison between ``model 1'' (orange) and ``model 2'' (blue) as spins in layers 3, 4, 7, and 8 deviate from the $\delta=0^{\circ}$ axis/the applied field direction. The other spins are either $\pm 68^{\circ}$ from the applied field direction, as determined by the Rietveld refined structures. Two values for $\left| J_2/J_1 \right|$ are reported, where solid lines represent $\left| J_2/J_1 \right| = 0.56$ (full model), and dashed lines represent $\left| J_2/J_1 \right| = 0.36$ (reduced model). In both cases, the energy increases linearly as the spins deviate from the applied field direction for ``model 2,'' whereas the total energy initially decreases for ``model 1.''}
\label{fig:mod1vmod2}
\end{figure}

The low temperature, high-field fan-like phase forms only below $\approx 170$ K. Interestingly, this is also the lower-bound temperature for the observed THE, as thermal fluctuations are greatly reduced below this temperature, which is perhaps why the FL phase can find stability. This postulation comes from the theory that the THE mechanism is a result of chiral fluctuations stabilized by thermal fluctuations. As discussed briefly in the Results section, the two fan-like models are almost indistinguishable, especially in the limit that $\delta \rightarrow 0$ in model 1 (detailed magnetic structure factor calculations for both models are given in the Appendix). Model 1 can be justified as being more likely the correct structure, though. Fig.~\ref{fig:mod1vmod2} shows the energy for each model as the spins in layers 3, 4, 7, and 8 deviate from $\delta=0^{\circ}$. Because the angles for spins 1, 2, 5, and 6 were found to be the same within error for both models, they were kept fixed at $\pm 68^{\circ}$ degrees for the calculations (as determined by the Rietveld refinement), which can be described by the equations,

\begin{align}
Model\; 1: & E = \left| J_2/J_1 \right|\cos(68^{\circ} + \delta ) - \cos (2 \delta ) \\
Model\; 2: & E = \left| J_2/J_1 \right|\cos(68^{\circ} - \delta ) - \cos (0^{\circ}),
\end{align}

where $J_1$ is the ferromagnetic exchange between layers within the nuclear unit cell, and $J_2$ is the antiferromagnetic exchange between layers on either side of the nuclear unit cell boundary, and the energy is in units of $J_1$. Only nearest neighbor interlayer exchange has been included in the calculation. The first term is the energy gain that results from spins connected by the $J_2$ exchange not being antiparallel. The second term is the reduction in energy due to the relative alignment between spins 3 and 4 (or 7 and 8), and it can be seen the the maximum reduction in energy is realized when the spins are ferromagnetically coupled. 

Although the zero-field magnetic structure in YMn$_6$Sn$_6$ has been solved for quite some time, \cite{VENTURINI1996102} polarized neutron diffraction has not been performed until now, and provides some much needed new information in the study of this intriguing material. For example, helical and spin-density wave magnetic structures can often be difficult to distinguish from one another using unpolarized neutrons in a single-crystal diffraction experiment. Even the addition of polarized neutrons with uniaxial polarization analysis may prove unhelpful depending on the scattering geometry with respect to the magnetic structure or if multiple helical domains are evenly populated. Here, polarized neutron diffraction results were able to show that the magnetic structure is helical and that the chiral domains are not evenly populated.  

In a uniaxial polarized neutron experiment, there are four neutron scattering cross-sections: $I^{++}$, $I^{+-}$, $I^{-+}$, and $I^{--}$. Nuclear coherent scattering never causes the reversal of the spin and hence is only observed in the $I^{++}$ and $I^{--}$ cross-sections. When the scattering vector, \textbf{Q}, is parallel to the neutron polarization, \textbf{P}, all nuclear scattering is in the non-spin-flip (NSF) channels, $I^{++}$ or $I^{--}$, and all magnetic scattering is in the spin-flip channels, $I^{+-}$ or $I^{-+}$ and hence can be distinguished unambiguously. Following the polarization analysis theory in Ref.~\cite{williams1988polarized}, the scattering intensities $I^{\pm \mp}$ are proportional to the spin-dependent cross-sections,

\begin{widetext}
\begin{equation}\label{eqn:xsections}
(d\sigma /d\Omega)_{\pm \mp} = \sum_{i,j}\exp[i\bm{\kappa}\cdot (\bm{r}_i - \bm{r}_j)]p_ip_j^* [\bm{S}_{\perp i} \cdot \bm{S}_{\perp j} \mp \sqrt{(-1)}\tilde{Z} \cdot (\bm{S}_{\perp i} \times \bm{S}_{\perp j}^*)],
\end{equation}
\end{widetext}

where the sum is over all magnetic atoms in the unit cell, $p=(\left| r_0 \right| /2)gf(Q)$ ($r_0$ is the neutron magnetic moment multiplied by the classical electron radius, $g$ is the Lande factor and $f(Q)$ is the magnetic form factor), $\bm{S}_i$ are the magnetic moment vectors, and $\tilde{Z}$ is a unit vector in the direction of the incoming neutron polarization.
The last term in Eqn.~\ref{eqn:xsections} is null for spin density waves and other collinear structures. In fact, there is no way to obtain unequal $I^{+-}$ and $I^{-+}$ intensities on a magnetic Bragg peak without imaginary components in the basis vectors, which result in a spiral-type structure. Typically, for a centrosymmetric crystal, chiral domains will be present in equal populations, because there is no energetic reason to favor one over the other. The scattering from the different domains then would result in equal $I^{+-}$ and $I^{-+}$ intensities. This is in contrast to single domain chiral crystals, where the sense (or sign) of the Dzyaloshinskii-Moriya interaction, if present, will pick out a single chiral domain, resulting in unequal $I^{+-}$ and $I^{-+}$ intensities. 

The results of the polarized neutron study point towards the ability to manipulate or switch the chirality. Control of magnetic properties with an electric current or electric field has been well-documented in multiferroics, materials exhibiting the magnetoelectric effect, and materials with broken inversion crystal symmetry. What can result in unequal spin-flip channel populations for centrosymmetric crystals is some external force to pick out a favorable chiral domain. For example, the simultaneous application of a magnetic field and electric current density was shown to control the chirality in MnP via spin transfer torque \cite{jiang2020electric}. The chiral inequality for the YMn$_6$Sn$_6$ sample used in this study was surprising because no such external perturbation was intentionally applied. Since the established crystal structure is achiral, this observation requires additional breaking of the $z\rightarrow -z$ mirror symmetry. For instance, this symmetry breaking may be effected through a particular defect ordering, or through asymmetric surface termination. In any event, this strong chirality not warranted by the underlying crystal structure is very interesting and deserves further investigation.    

\section{Summary}
A magnetic field applied in the $ab$-plane of YMn$_6$Sn$_6$ leads to an extensive field-temperature phase diagram. This is owed to the delicate balance of competing interplane exchange interactions between the magnetic kagome lattice layers. The neutron diffraction results presented here solve the magnetic structures for two of the previously identified phases. These are the commensurate canted antiferromagnet (CAF), appearing around room-temperature and low fields, and the commensurate fan-like (FL) structure, which appears at low-temperatures and high-fields. Our study also revealed an additional incommensurate magnetic structure exists between the FL and forced ferromagnetic (FF) phases, which explains the ``Phase I'' region previously identified in ac susceptibility measurements. Two incommensurate wavevectors appear throughout many regions of the phase diagram, including at zero-field, where both magnetic structures are the double-flat-spiral, but with slightly differing periodicities. Via our high resolution synchrotron powder diffraction measurements, we were able to show that the presence of the two wavevectors is likely an intrinsic feature of YMn$_6$Sn$_6$. Polarized neutron diffraction measurements showed that the zero-field incommensurate magnetic structures have preferential, but opposite, chiralities, which is a phenomena usually reserved for lattices with broken inversion symmetry. 

\section{Acknowledgments}
Synthesis and characterization work (N.J.G.) were supported by the U.S. Department of Energy, Office of Science, Basic Energy Sciences, Materials Science and Engineering Division. I.I.M.\ acknowledges support from the U.S. Department of Energy through the grant \#DE-SC0021089. Use of the Advanced Photon Source at Argonne National Laboratory was supported by the U. S. Department of Energy, Office of Science, Office of Basic Energy Sciences, under Contract No. DE-AC02-06CH11357. The identification of any commercial product or trade name does not imply endorsement or recommendation by the National Institute of Standards and Technology.

\appendix*
\section{structure factor calculations}
\begin{figure}[t]
\includegraphics[scale=0.45]{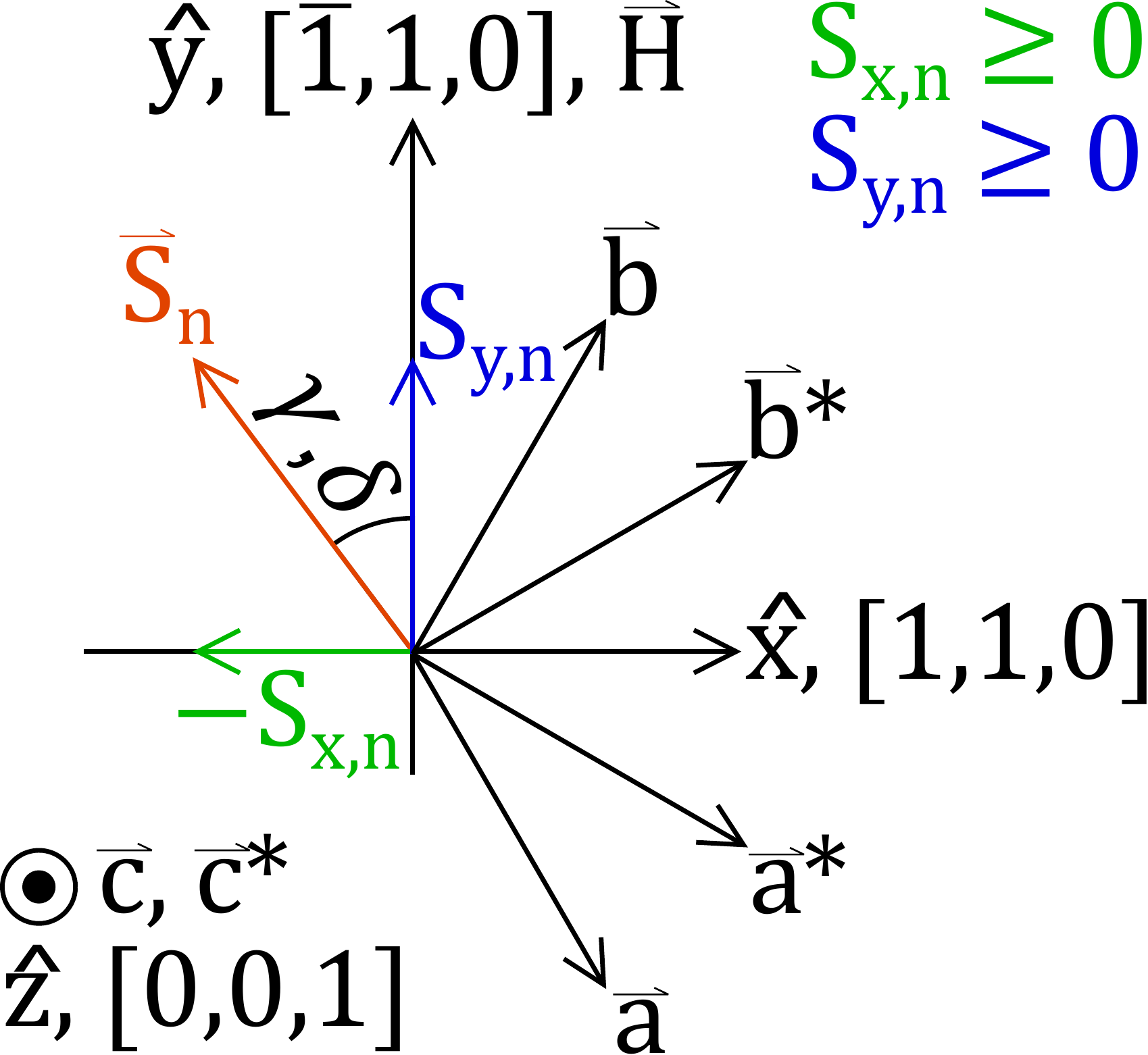}
\caption{The coordinate system used in calculating the magnetic structure factors, where $\widehat{\mathrm{x}}$, $\widehat{\mathrm{y}}$, and $\widehat{\mathrm{z}}$ are unit vectors which define a right-handed Cartesian coordinate system. The vector, $\vec{\mathrm{S}}_n$, is the spin magnitude and direction for Mn moments in layer $n$, and $\gamma$ and $\delta$ are the values defined by the angle that $\vec{\mathrm{S}}_n$ makes with the applied field direction, $\vec{\mathrm{H}}$. }
\label{fig:supp_SFgeom}
\end{figure}

To elucidate any differences between the low-temperature, high-field models (model 1 and model 2), we calculated the magnetic structure factor, $\overline{\mathrm{F}}_{\mathrm{M}}$, for various reflections in the $(H,H,L)$ scattering plane. The geometry for calculating the models is shown in Fig.~\ref{fig:supp_SFgeom}, where $n$ ($n=1,2,3,...,8$) refers to a layer of Mn moments in the magnetic unit cell ($a$, $b$, $4c$), and the $n=1$ layer is that which is closest to the $c$-axis origin (above that layer is $n=2$, etc.). The vectors, $ \overline{\mathrm{S}}_n$, give the magnitude and direction of spins in layer $n$. All spins can be defined by $\overline{\mathrm{S}}_1$ and $\overline{\mathrm{S}}_3$, which define the angles $\gamma$ and $\delta$, respectively. Because neutrons are only sensitive to the component of spin which is perpendicular to the scattering vector, $\overline{\mathrm{S}} _{\perp}$, we must also define the scattering vector in the same coordinate system as $\overline{\mathrm{Q}} = Q\widehat{\mathrm{e}}$, where $\widehat{\mathrm{e}} = e_x \widehat{\mathrm{x}} + e_y \widehat{\mathrm{y}} + e_z \widehat{\mathrm{z}}$ is a unit vector parallel to the scattering vector. In the $(H,H,L)$ scattering geometry, $e_y$ is always zero, and we have assumed moments are in the $ab$-plane so that $S_{z,n}=0$. $\overline{\mathrm{S}} _{\perp}$ can then be written as,

\begin{align}
\begin{split}
\overline{\mathrm{S}} _{\perp ,n} & = \overline{\mathrm{S}} _n - \widehat{\mathrm{e}} \left( \widehat{\mathrm{e}} \cdot \overline{\mathrm{S}} _n \right) \\
& = S_{x,n} \left( 1 - e_x ^2 \right) \widehat{\mathrm{x}} + S_{y,n} \widehat{\mathrm{y}} - e_x e_z S_{x,n} \widehat{\mathrm{z}}.
\label{eqn:Sperp}
\end{split}
\end{align}

In the following structure factor calculations, $L$ is with respect to the magnetic unit cell (divide by four to get the equivalent reflection in the nuclear unit cell), and the prefactor is $p=(\left| r_0 \right| /2)g f(Q)$, where $r_0$ is the neutron magnetic moment multiplied by the classical electron radius, $g$ is the Lande factor and $f(Q)$ is the magnetic form factor. The magnetic structure factor for all $(H, H, 0)$ peaks for both models is,

\begin{equation}
\left| \overline{\mathrm{F}}_{\mathrm{M}} \right| _{H,H,0} ^2 = 144 \left| p \right| ^2 \left( S_{y,1} + S_{y,3} \right) ^2.
\label{eqn:HH0}
\end{equation}

The magnetic structure factor for $(H,H,L)$ peaks with $L=2m$ (where $m=1,2,3,...$, and $ \theta (x)$ is the Heaviside step function) for both models is,

\begin{align}
\begin{split}
&\left| \overline{\mathrm{F}}_{\mathrm{M}} \right| _{H,H,L=2m} ^2 = 144 \left| p \right| ^2 [ \theta \left( \left( -1 \right) ^m \right) - \left( -1^m \right) \\
& \times \sin ^2 \left( 4 \pi m r_{c,1} \right) ] \left( S_{y,1} + \left( -1 \right) ^m S_{y,3} \right) ^2.
\label{eqn:HHL_2m}
\end{split}
\end{align}

The magnetic structure factor for $(H,H,L)$ peaks with $L=2m+1$ for \textbf{model 1} is,
\begin{align}
\begin{split}
& \left| \overline{\mathrm{F}}_{\mathrm{M}} \right| _{H,H,L=2m+1} ^2 =  72 \left| p \right| ^2 \left[ \left( 1 - e_x ^2 \right) ^2 + e_x ^2 e_z ^2 \right] \\
& \times [ S_{x,1} ^2 + S_{x,3} ^2 - 2S_{x,1} S_{x,3} \cos \left( 4 \pi \left( 2m+1 \right) r_{c,1} \right) \\
& + \left( -1 \right) ^m \left( S_{x,1} ^2 - S_{x,3} ^2 \right ) \sin \left( 4 \pi \left( 2m+1 \right) r_{c,1} \right) ].
\label{eqn:HHL_2mp1_mod1}
\end{split}
\end{align}

The magnetic structure factor for $(H,H,L)$ peaks with $L=2m+1$ for \textbf{model 2} is,
\begin{align}
\begin{split}
& \left| \overline{\mathrm{F}}_{\mathrm{M}} \right| _{H,H,L=2m+1} ^2 = 72 \left| p \right| ^2 \left[ \left( 1 - e_x ^2 \right) ^2 + e_x ^2 e_z ^2 \right] \\
& \times \left( S_{x,1} ^2 + S_{x,3} ^2 \right) \left[ 1 + \left( -1 \right) ^m \sin \left( 4 \pi \left( 2m+1 \right) r_{c,1} \right) \right].
\label{eqn:HHL_2mp1_mod2}
\end{split}
\end{align}

\begin{figure}[b]
\includegraphics[scale=0.28]{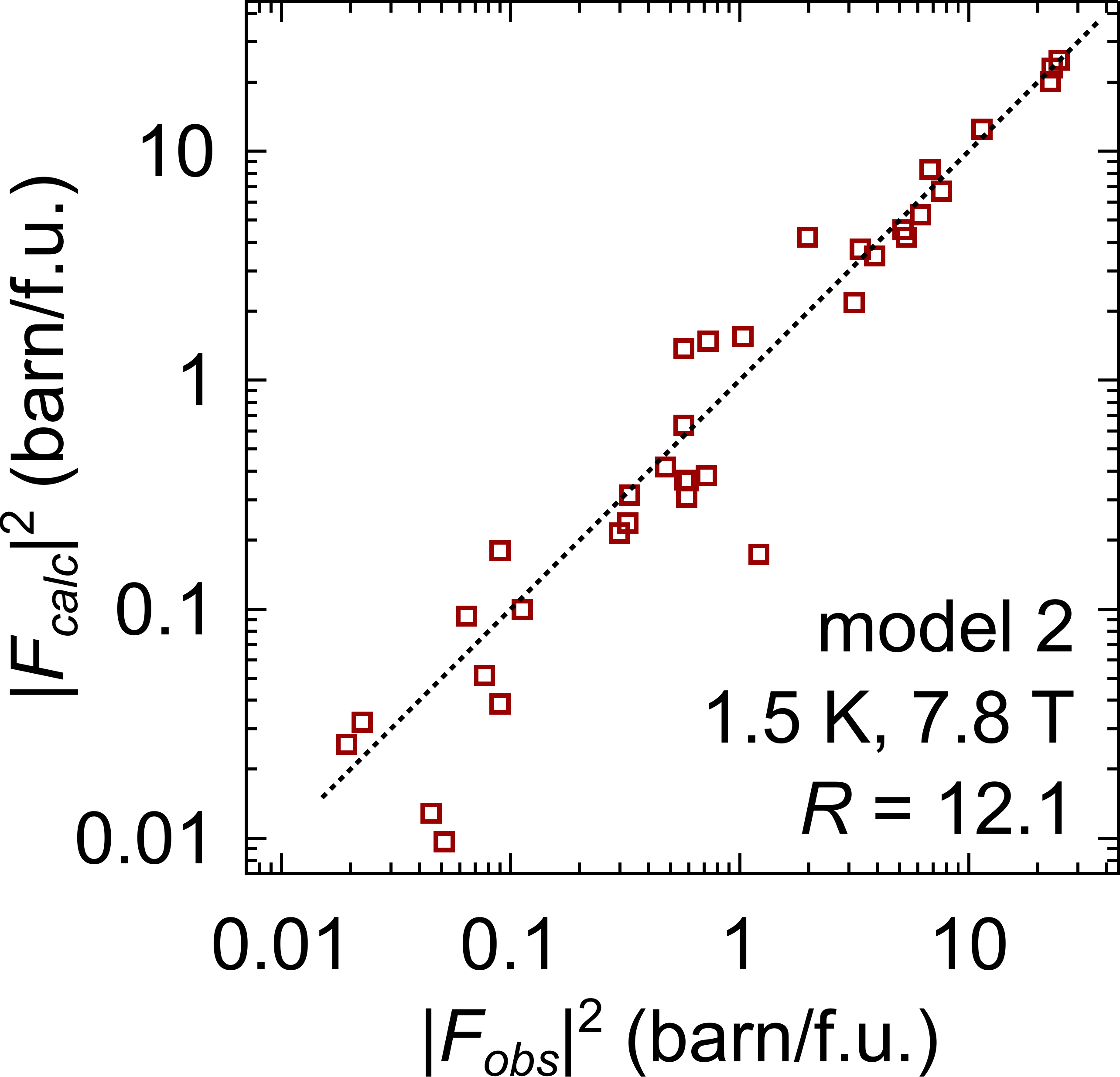}
\caption{The observed versus calculated magnetic structure factor squared results for the magnetic 'model 2' for the low temperature high-field phase. The calculated result was obtained via Rietveld refinement for the magnetic intensity only. Magnetic Bragg peaks have wavevectors, $(0, 0, 0)$, $(0, 0, 0.25)$, and $(0, 0, 0.5)$. Data were taken at 1.5 K and 7.8 T in the $(H, H, L)$ scattering plane.}
\label{fig:LTHF_m2}
\end{figure}

The structure factors for both models have the same dependency on the $S_y$ components, and only the $S_y$ components, for $(H,H,0)$ and $(H,H,L=2m)$ peaks. These include the peaks coincident with the nuclear Bragg peaks and magnetic Bragg peaks at $L=0.5$ of the nuclear unit cell. The $S_y$ components are those which are along the applied field direction. The models differ for the structure factors with $L=0.25$ or $L=0.75$ of the nuclear unit cell. This can be seen by Eqns.~\ref{eqn:HHL_2mp1_mod1} and \ref{eqn:HHL_2mp1_mod2}. The structure factors for these peaks are only dependent on the $S_x$ components of the spins, and it can be seen why the model 1 and model 2 refinements give practically the same goodness of fit: as $S_{x,3} \rightarrow 0$, which equivalently means $\delta \rightarrow 0$, Eqns.~\ref{eqn:HHL_2mp1_mod1} and \ref{eqn:HHL_2mp1_mod2} converge. The refined $\delta$ for model 1 is zero within error, and $\delta$ for model 2 is $12 ^{\circ}$ with a large error of $\pm 6 ^{\circ}$. Both the moment size and angle, $\gamma$ (defined by $\vec{\mathrm{S}}_1$), are the same within error for model 1 and model 2. The model 2 refinement results are shown in Fig.~\ref{fig:LTHF_m2}

\end{document}